\documentstyle[11pt]{article}

\newcommand{\beq}{\begin{equation}}
\newcommand{\eeq}{\end{equation}}
\newcommand{\beqa}{\begin{eqnarray}}
\newcommand{\eeqa}{\end{eqnarray}}

\newcommand{\NPB}[1]{{\it Nucl. Phys.}\ {\bf B{#1}}}
\newcommand{\PLB}[1]{{\it Phys. Lett.}\ {\bf B{#1}}}
\newcommand{\PRD}[1]{{\it Phys. Rev.}\ {\bf D{#1}}}

\begin{document}

\begin{titlepage}
\def\thepage {}        
\title{
Isospin Breaking and Fine Tuning in Top-Color\\
Assisted Technicolor}

\author{
R.S. Chivukula\thanks{e-mail addresses: sekhar@abel.bu.edu,
dobrescu@buphy.bu.edu, terning@calvin.bu.edu},
B.A. Dobrescu, and J. Terning\\
Department of Physics, Boston University, \\
590 Commonwealth Ave., Boston MA  02215}

\maketitle

\bigskip
\begin{picture}(0,0)(0,0)
\put(295,250){BUHEP-95-8}
\put(295,235){hep-ph/9503203 }
\end{picture}
\vspace{12pt}
\date{}

\begin{abstract}

Recently, Hill has proposed a model in which new, potentially
low-energy, top-color interactions produce a top-condensate ({\em a
la} Nambu---Jona-Lasinio) and accommodate a heavy top quark, while
technicolor is responsible for producing the $W$ and $Z$ masses.  Here
we argue that isospin breaking gauge interactions, which are necessary
in order to split the top and bottom quark masses, are likely to
couple to technifermions. In this case they produce a significant
shift in the $W$ and $Z$ masses (i.e.~contribute to $\Delta \rho_* =
\alpha T$) if the scale of the new interactions is near 1 TeV.  In
order to satisfy experimental constraints on $\Delta
\rho_*$, we find that either the effective top quark coupling or the
top-color coupling must be adjusted to 1\%. Independent of the
couplings of the technifermions, we show that the isospin-splitting of
the top and bottom quarks implies that the top-color gauge bosons must
have masses larger than about 1.4 TeV.  Our analysis can also be
applied to strong extended technicolor (ETC) models that produce the
top-bottom splitting via isospin breaking ETC interactions.
\pagestyle{empty}
\end{abstract}
\end{titlepage}
\section{Introduction}
\label{sec:intro}
\setcounter{equation}{0}

Recently, Hill has combined aspects of two different approaches to
dynamical electroweak symmetry breaking into a model which he refers
to as top-color assisted technicolor (${\rm TC}^2$) \cite{top-color}.
In this model a top-condensate is driven by the combination of a
strong isospin-symmetric top-color interaction and an additional
(either weak or strong) isospin-breaking $U(1)$ interaction which
couple only to the third generation quarks. He argues that the extreme
fine-tuning that was required in pure top-condensate models can be
done away with if the scales of the critical top-color and $U(1)$
interactions are brought down to a TeV.  Given a top quark mass of
around 175 GeV, such top-color interactions would produce masses
for the $W$ and $Z$ that are far too small, hence there must be further
strong interactions (technicolor) that are primarily responsible
for breaking electroweak symmetry.

At low-energies, the top-color and hypercharge interactions of the
third generation quarks may be approximated by four-fermion operators
\cite{top-color}
\beq
{\cal L}_{4f} = -{{4\pi
\kappa_{tc}}\over{M^2}}\left[\overline{\psi}\gamma_\mu {{\lambda^a}\over{2}}
\psi \right]^2
-{{4\pi \kappa_1}\over{M^2}}\left[{1\over3}\overline{\psi_L}\gamma_\mu  \psi_L
+{4\over3}\overline{t_R}\gamma_\mu  t_R
-{2\over3}\overline{b_R}\gamma_\mu  b_R
\right]^2~,
\label{L4t}
\eeq
where $\psi$ represents the top-bottom doublet, $\kappa_{tc}$ and
$\kappa_1$ are related respectively to the top-color and $U(1)$
gauge-couplings squared, and where (for convenience) we have assumed
that the top-color and $U(1)$ gauge-boson masses are comparable and of
order $M$. The first term in equation (\ref{L4t}) arises from the
exchange of top-color gauge bosons, while the second term arises from
the exchange of the new $U(1)$ hypercharge gauge boson which has
couplings proportional to the ordinary hypercharge couplings.  In
order to produce a large top quark mass without giving rise to a
correspondingly large bottom quark mass, the combination of the
top-color and extra hypercharge interactions are assumed to be
critical in the case of the top quark but not the bottom quark.  The
criticality condition\footnote{The criticality condition given in
ref.~\cite{top-color} is inconsistent with the interaction Lagrangian
in equation (\ref{L4t}), i.e. equation (4) of ref.~\cite{top-color}.
In the large-$N_c$ limit, $\kappa_c = \pi/3$.}  for top quark
condensation in this model is then:
\beq
\kappa^t_{eff} = \kappa_{tc} +{1\over3}\kappa_1 >
\kappa_c = {{3\pi}\over{8}} >
\kappa^b_{eff}=\kappa_{tc} -{1\over 6}\kappa_1~.
\label{kc}
\eeq

In order that the dynamical scale of the theory, $M$, be of order
1 TeV while the top quark mass is of order 175 GeV, some degree
of fine-tuning is required. We can quantify the amount of fine-tuning
involved to keep the top quark lighter than $M$ in terms of
\beq
{\Delta\kappa_{eff}\over\kappa_c} = {{\kappa^t_{eff}-\kappa_c}
\over \kappa_c}~.
\label{dkeff}
\eeq
The gap-equation for the Nambu--Jona-Lasinio model implies that
\beq
{\Delta\kappa_{eff}\over\kappa_c} = {{{m^2_t\over M^2}\log{M^2\over
m^2_t}} \over {1-{{m^2_t\over M^2}\log{M^2\over m^2_t}}}}~.
\label{dkefft}
\eeq
For $m_t$=175 GeV and $M$=1 TeV, we find that
\beq
{\Delta\kappa_{eff}\over\kappa_c} \approx 12\%~.
\label{dkeffn}
\eeq
As emphasized by Hill, since $M$ is of order 1 TeV, it is possible
to accommodate a top quark mass of 175 GeV without excessively fine-tuning
$\kappa^t_{eff}$.

In this paper we will argue that the extra $U(1)$ interaction in the
${\rm TC}^2$ model is likely to also couple to technifermions in an
isospin-breaking fashion. We will show that this produces a
potentially large contribution to\footnote{ We are distinguishing here
between isospin breaking that is present in the standard model,
$\Delta \rho$, and isospin breaking from new physics, $\Delta \rho_*
$.} $\Delta \rho_* \equiv \alpha T$, \cite{Drho,T} even if one assumes
that there is {\em no} isospin breaking in the technifermion masses.
Thus the isospin breaking $U(1)$ interaction in such models is likely
to have to be weak. If this $U(1)$ interaction is weak, however,
equation (\ref{kc}) implies fine-tuning of the top-color coupling
$\kappa_{tc}$. We will show that either the tuning required to keep
the top quark light compared to $M$ (equation (\ref{dkefft})), or that
required for $\kappa_{tc}$ because the $U(1)$ coupling is weak
\beq
{\Delta\kappa_{tc}\over\kappa_c} = \left|{{\kappa_{tc}-\kappa_c}
\over \kappa_c}\right| \le
{1 \over 3} {\kappa_1\over \kappa_c}~,
\label{dktc}
\eeq
must be finer than 1\%.

We further show that, independent of any assumption about the $U(1)$
couplings of the technifermions, the new top-color interactions for
the top quark are themselves phenomenologically unacceptable at low
scales.  In particular, assuming that the top-bottom splitting is the
{\em only} source of weak isospin breaking, if the top-color
interactions are strong enough to produce a top-condensate, then the
bound on $\Delta \rho_*$ requires that the mass of the top-color boson
be larger than about 1.4 TeV. This bound is sufficient to rule out the
possibility that these new gauge bosons produce a significant
enhancement \cite{topprod} of the top production cross section at
FNAL.

In the next section we study the contributions to $\Delta \rho_*$ from
technifermion loops.  Next we calculate the contribution to $\Delta
\rho_*$ from the new strong top interactions employed in the ${\rm
TC}^2$ model.  Finally we compare these results with experimental
data, and present out conclusions.

\section{Technifermion Loops}
\label{sec:technifermion}
\setcounter{equation}{0}

Since the ${\rm TC}^2$ model relies largely on technicolor to provide
masses for the $W$ and $Z$, the technifermions in the same weak
doublets must be approximately degenerate.  It is difficult to achieve
this degeneracy in any technicolor model, since the technifermions
typically have some couplings to the top and bottom quarks, and thus the
required isospin breaking dynamics can feed into technifermion mass
splittings, and produce a contribution to $\Delta \rho_*$
\cite{isospin1}.  Even though such a contribution is potentially
dangerous, the effect is extremely difficult to calculate, and is
quite model dependent, so we will not attempt to calculate it here.

However, even if the technifermions are degenerate, there will be
additional significant (and positive) contributions to $\Delta \rho_*$
\cite{isospin2} if the additional $U(1)$ interaction couples to the
right-handed technifermions in a manner that violates custodial
symmetry.  Note that some of the technifermions must couple to the
additional $U(1)$: part of the top quark mass must come from ETC
interactions in order to prevent the appearance of an extra set of
light isotriplet Goldstone bosons.  If the ETC interactions commute
with $SU(2)_W$ and if the bottom quark also has a mass coming from the
ETC interactions, then (since the top and bottom quarks must have
differing $U(1)$ charges to allow for their different masses) the
different right-handed technifermions to which top and bottom quarks
couple must have different (and hence isospin-violating) charges.

In order to avoid these contributions to $\Delta\rho_*$, one must have
custodially invariant $U(1)$ couplings to the technifermions.  This
will be difficult to accomplish while still evading the manifold
anomaly cancellation constraints. (In order to do so, one will need to
introduce exotic charges for the technifermions (cf. Bouchiat,
Iliopoulos, and Meyer \cite{bim})). Furthermore, one must also allow
for the third generation, which carries the new $U(1)$, to mix with
the first and second generations, which carry custodial-symmetry
violating hypercharge.

In this section we show that, if the additional $U(1)$ interactions
violate custodial symmetry, the $U(1)$ coupling will have to be quite
small to keep this contribution to $\Delta \rho_*$ small.  We will
illustrate this in the one-family technicolor \cite{onefam} model,
assuming that techniquarks and technileptons carry $U(1)$-charges
proportional to the hypercharge of the corresponding ordinary
fermion\footnote{Note that this choice is anomaly-free.}. We can
rewrite the effective $U(1)$ interaction of the technifermions as
\beq
{\cal L}_{4T1} =-{{4\pi
\kappa_1}\over{M^2}}\left[{1\over3}\overline{\Psi}\gamma_\mu  \Psi
+\overline{\Psi}_R\gamma_\mu  \sigma^3 \Psi_R
-\overline{L} \gamma_\mu L
+ \overline{L}_R \gamma_\mu\sigma^3 L_R
\right]^2~,
\label{L4T}
\eeq
where $\Psi$ and $L$ are the techniquark and technilepton
doublets respectively.
Only the chiral piece of the
four-fermion interactions will contribute to $\Delta \rho_*$.

The shift in the ratio of (zero-momentum) $W$ and $Z$ masses from physics
beyond
the standard model  is given by \cite{T}:
\beq
\Delta \rho_* \equiv \alpha T =  {{e^2}\over{s^2 c^2 M_Z^2}}
\left( \Pi_{11}(q^2=0)-\Pi_{33}(q^2=0)\right)~,
\eeq
where  $\Pi_{11}(q^2)$ and $\Pi_{33}(q^2)$ are the contributions from
new physics to the coefficients of $i g_{\mu\nu}$ in the
$W$ and $Z$ vacuum polarizations, with the gauge couplings factored out.
Following the standard notation, $e$ is the electromagnetic gauge coupling,
$s$ and $c$ are the sine and cosine of the weak angle, and $M_Z$ is the
mass of the $Z$ gauge boson.

If the technifermions are degenerate, then the only contribution
\cite{isospin2} to $\Delta \rho_*$ will come from the square of the
right-handed interaction in equation (\ref{L4T}).  Thus the
technifermion contribution to the vacuum polarization $\Pi_{33}(0)$ is
just the product of two vacuum polarizations (see figure 1):
\beqa
\Pi_{33}^{\rm T}(0) &=& -{{8 \pi \kappa_1}\over{M^2}} \left({1\over
2}Tr(\sigma^3 \sigma^3)
\,\Pi_{LR}^{\rm T}(0)\right)^2 \nonumber \\
&=&-{{2\pi \kappa_1}\over{M^2}} (N_c+1)^2 f^4~,
\eeqa
where the technipion decay constant, $f$, is related to the electroweak
symmetry breaking scale, $v=246$ GeV, by:
\beq
v^2 = (N_c+1) f^2~.
\eeq
Thus the contribution to $\Delta \rho_*$ from degenerate technifermions is:
\beqa
\Delta \rho_{*}^{\rm T}&=&8 \pi  \kappa_1 {{v^2}\over{M^2}} \nonumber \\
&=&152\% \ \kappa_1 \left({{1\ {\rm TeV}}\over{M}}\right)^2~.
\label{rhoT}
\eeqa
We see immediately that, if $M$ is of order 1 TeV, $\kappa_1$ must
be extremely small.  We will see precisely how small in section
\ref{sec:data}.

\section{Top Quark Loops}
\label{sec:topcolor}
\setcounter{equation}{0}

Introducing new interactions for the top and bottom quarks will also
generate additional contributions to $\Delta \rho_*$.  In this
section, we will focus on the contributions to $\Delta \rho_*$ coming
from loops of top  and bottom quarks.

The contribution to $\Delta \rho_*$ from one top-color gauge boson
exchange (expressed in terms of vacuum polarizations for left-handed
currents) is:
\beq
\Delta \rho_{*}^{\rm tc} = {{e^2}\over{s^2 c^2 M_Z^2}}\left( {1\over2}
\Pi_{LL}(q^2=0,m_t,m_b)-{1\over 4}\Pi_{LL}(q^2=0,m_t,m_t)\right)~,
\eeq
where we have made explicit the dependence on the fermions (i.e.~top
and bottom) propagating in the two-loop graph\footnote{This
effect is suppressed by $f_t^2/M^2$ and is not included in the leading
order (in $1/M^2$) analysis of the gap equation.} (see figure 2).
We are neglecting terms suppressed by $(m_b/m_t)^2$.  In the
Nambu--Jona-Lasinio approximation we are considering, the
two-loop vacuum polarization calculation reduces to the product of
two one-loop vacuum polarizations.  Keeping only (universal) logarithmically
divergent terms in each of these one-loop pieces we find:
\beqa
\Pi_{LL}(q^2=0,m_t,m_t)&=&-{{64 \pi \kappa_{tc}}\over{9}}
{{f_t^4}\over{M^2}}~,\\
\Pi_{LL}(q^2=0,m_t,m_b)&=&-{{8\pi \kappa_{tc}}\over{9}}
{{f_t^4}\over{M^2}}~,
\eeqa
where
\beq
f_t^2 \equiv    {{N_c }\over{8\pi^2}}\, m_t^2
\log\left({{M^2}\over{m_t^2}}\right)~.
\eeq
Note that, with this definition, in the ${\rm TC}^2$ model  the one-top-loop
piece of the
$Z$ self energy contributes
\beq
{{e^2}\over{4 s^2 c^2}}f_t^2
\eeq
to $M_Z^2$.

Putting everything together we find the top-color exchange contribution to be:
\beq
\Delta \rho_{*}^{\rm tc} = {{4 \pi e^2 \kappa_{tc}}\over{3s^2 c^2 }}
{{f_t^4}\over{M_Z^2 M^2}}~.
\label{rhotc}\eeq
Since $\kappa_1$ must be small, equation (\ref{kc}) implies
that $\kappa_{tc}$ must be close
to $3\pi/8$.
Taking  $\kappa_{tc} \approx  3\pi/8$, $m_t= 175$ GeV and $M =1$ TeV, we have:
\beq
\Delta \rho_{*}^{\rm tc} \approx 0.53\%~,
\eeq
which is about one half of the one-loop top quark contribution to
$\Delta \rho$ in the standard model. Note that this contribution is
nonzero, even though the top-color interactions themselves are {\em
isospin-symmetric} -- isospin violation is present in the form of the
$t$-$b$ mass splitting. We will see in the next section that
experimental data constrain $\Delta \rho_*$ to be less than this
value.

There is also a contribution to $\Delta \rho_*$ from
the $U(1)$ interaction  with one top quark loop and one technifermion
loop.  Assuming (again) that the techniquarks are
degenerate, this contribution is:
\beq
\Delta \rho_{*}^{\rm T, t} = 8 \pi  \kappa_1 {{f_t^2}\over{M^2}}~.
\label{rhoTt}
\eeq
Thus for $m_t = 175$ GeV and $M = 1$ TeV:
\beq
\Delta \rho_{*}^{\rm T, t} = 10.2\%\ \kappa_1~.
\label{rhoTtn}
\eeq
The contribution from one $U(1)$ gauge boson exchange
with two quark loops is even smaller.

\section{Comparison with Data}
\label{sec:data}
\setcounter{equation}{0}
In order to extract the $S$ and $T$ parameters, we have performed
(using the procedure described in ref. \cite{fit}) a
global fit to precision electroweak data.  We have used the most
precise measurement \cite{lattice} of $\alpha_s(M_Z^2) = 0.115 \pm 0.002$.
In figure 3 we show the ellipse in the $S-T$ plane which projects onto the
95\% confidence interval on the $T$ axis.  This gives the range  $-0.25 < T <
0.52$
($-0.19\% < \Delta \rho_* < 0.40\%$) as the 95\% confidence interval.
Using a larger value for $\alpha_s$ results in an even tighter bound on $T$.
For example, using $\alpha_s(M_Z^2) = 0.124$ (which is obtained from
a  global fit to precision electroweak  data \cite{langacker}) we find $-0.46 <
T < 0.30$
($-0.36\% < \Delta \rho_* < 0.23\%$).

If top-color exchange (with top and or bottom quarks in the loops,
equation (\ref{rhotc})) was the only contribution to $\Delta \rho_*$,
the mass of the top-color gauge boson would have to be larger than
about 1.4 TeV (using $\alpha_s(M_Z^2) = 0.124$ raises this bound to
2.2 TeV), in order for $\Delta \rho_*$ to remain acceptably small.
While this is still of order 1 TeV, the bound is sufficient to rule
out the possibility that these new gauge bosons produce a significant
enhancement \cite{topprod} of the top production cross section at
FNAL\footnote{Of course, the NJL calculations can only be taken as
an approximation to the calculation in the full top-color theory. However,
in order for top-color production to significantly enhance top
quark production we need $M$ to be of order 400 - 500 GeV \cite{lane}.}.

If we require that the total contribution to $\Delta \rho_*$
(the sum of equations (\ref{rhoT}), (\ref{rhotc}),
and (\ref{rhoTt})) is less than 0.40\%, we bound $\kappa_1$ from
above and $M$ from below. When $\kappa_1$ is small the
top-color coupling must be adjusted carefully (see equation
(\ref{dktc})). When $M$ is large, more fine-tuning is required to
keep the top quark light (equation (\ref{dkefft})).
For $M>$ 1.4 TeV, we find that either $\Delta\kappa_{tc}/
\kappa_c$ or $\Delta\kappa_{eff}/\kappa_c$ must be tuned to
less than 1.0\%.  This trade-off
in fine tunings is displayed in figure 4.  For the ``best"
case where both tunings are 1.0\%, $M=4.5$ TeV.
 (If we require $\Delta \rho_*$ be less than 0.23\%, we find that the
couplings must be adjusted to approximately 0.7\%
and $M=5.3$ TeV for the ``best" case scenario.)

\section{Conclusions}
\label{sec:concl}

We have found that the ${\rm TC}^2$ model can produce potentially
dangerous shifts in the $W$ and $Z$ masses.  If the new $U(1)$
interaction has custodial violating couplings to the technifermions,
the experimental limit on isospin breaking in the $W$ and $Z$ masses
forces the coupling of the $U(1)$ gauge boson to be small or the scale
of the new interactions to be large.  A small $U(1)$ coupling would
imply a certain amount of fine-tuning. In particular, for the
technifermions in the one-family model and $U(1)$ couplings
proportional to hypercharge, we find that either the effective top
quark coupling or the top-color coupling must be adjusted to an
accuracy of at least 1\%.  For generic gauge boson masses (i.e.~away
from $M=4.5$ TeV), the fine-tuning required is even more
severe.  Of course, with enough fine-tuning one could arbitrarily
increase the top-color scale and dispose of the problem of isospin
breaking, but the primary motivation of the model was to avoid
fine-tuning.

These bounds can be mitigated if the $U(1)$ couplings do not
violate custodial symmetry. We believe that constructing such
a model will be difficult, given that one must cancel all
gauge anomalies and allow for mixing between the third generation
with the first two.

Independent of whether the $U(1)$ couplings to technifermions violate
custodial symmetry, the contribution to $\Delta\rho_*$ from top-color
exchange implies that the top-color gauge boson mass must be greater
than approximately 1.4 TeV.

There are analogous effects in strong ETC models \cite{scal}, for a
recent example see ref.~\cite{wu}.  There, as well, the direct effect
on $\Delta \rho_*$ from ETC gauge boson exchange can be suppressed at
the expense of fine-tuning, while the lack of isospin symmetry in the
technifermion spectrum remains problematic.  The bound on the mass of
the top-color gauge boson that we obtained from the contribution of
top and bottom quark loops (discussed in section \ref{sec:topcolor})
can be directly taken over to the strong ETC case.

\centerline{\bf Acknowledgments}

We thank Dimitris Kominis for helpful conversations, Ken Lane for
helpful discussions and a careful reading of the manuscript,
and Chris Hill for useful comments.
R.S.C. acknowledges the support of  an NSF Presidential Young
Investigator Award, and a DOE Outstanding Junior Investigator Award.
{\em This work was supported in part by the National Science
Foundation under grant PHY-9057173, and by the Department of Energy under
grant DE-FG02-91ER40676.}


\newpage
\centerline{\bf Figure Captions}
\bigskip  \noindent
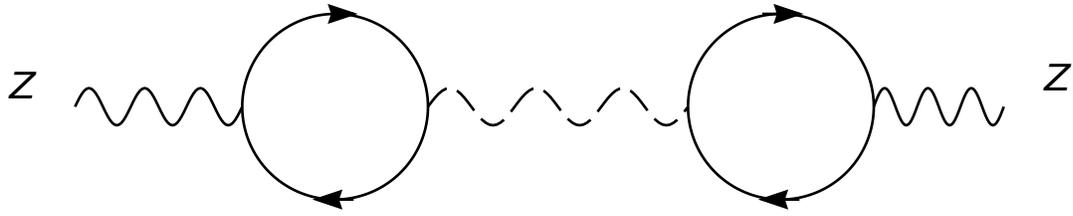
{\bf Figure 1.}  The $U(1)$ gauge boson contribution to the $Z$
vacuum polarization.
\bigskip \newline \noindent
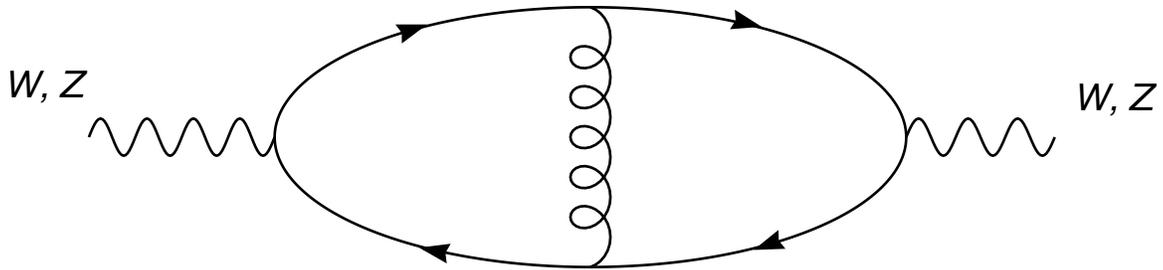
{\bf Figure 2.}  The top-color gauge boson contribution to the $W$ and $Z$
vacuum polarizations, with top and bottom quarks in the loop.
\bigskip \newline \noindent
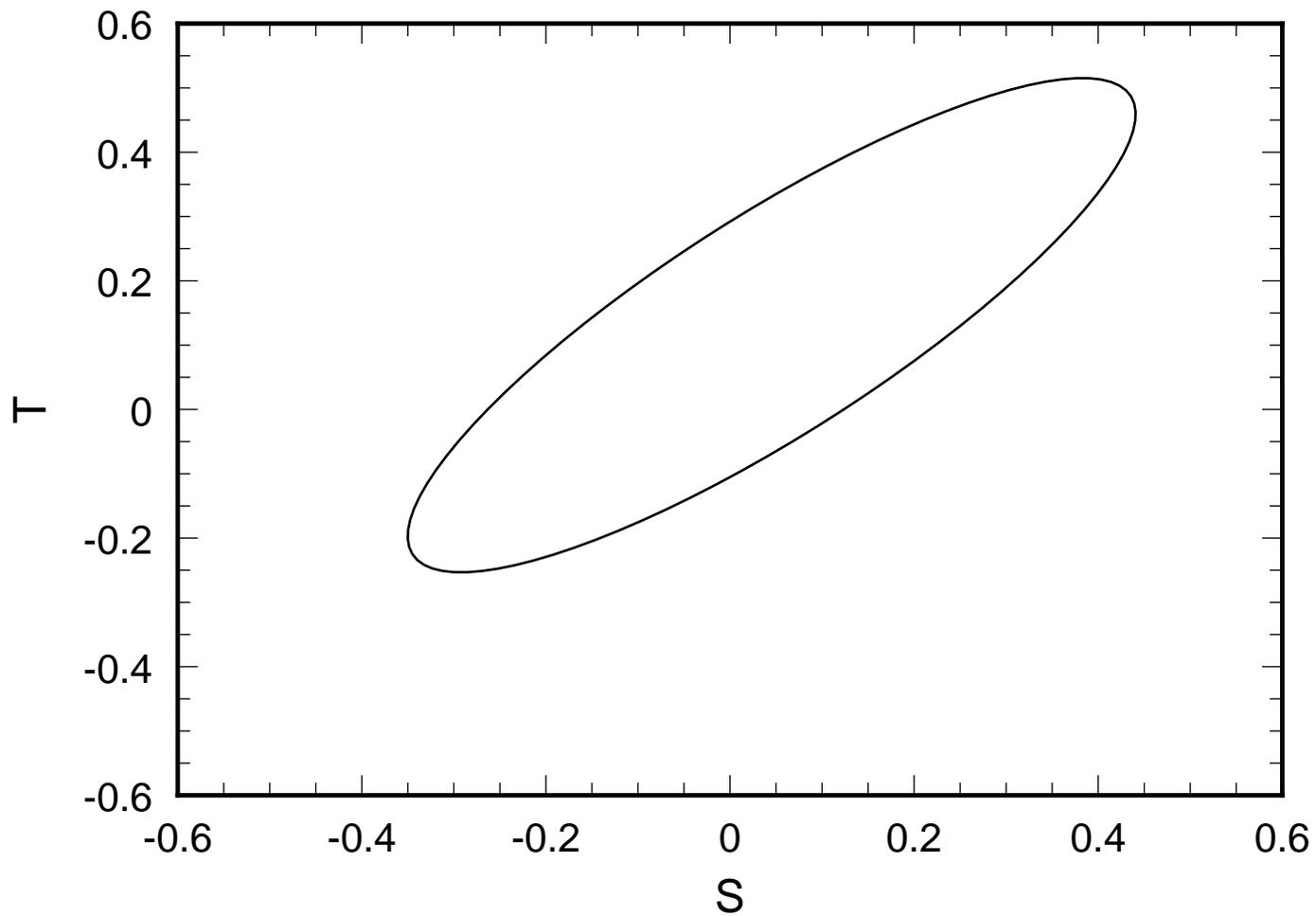
{\bf Figure 3.} The ellipse in the $S$-$T$ plane which projects onto the
95\% confidence range for $T$.
\bigskip \newline \noindent
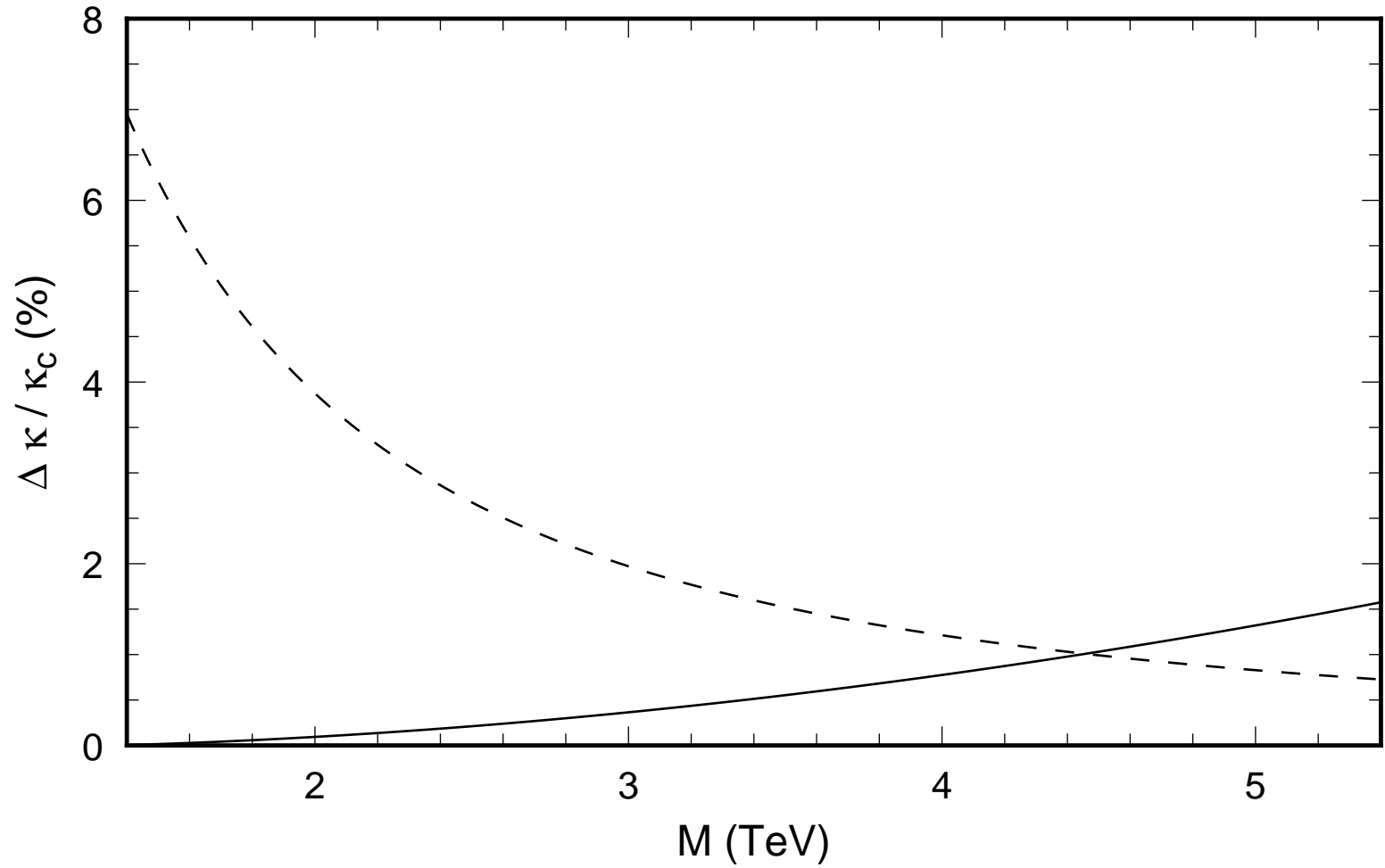
{\bf Figure 4.} The amount of fine-tuning required in the ${\rm TC}^2$ model.
The dashed line is the amount of fine-tuning in $\Delta\kappa_{eff}$ required
to
keep $m_t$
much lighter than $M$, see equation (\ref{dkefft}). The solid curve shows
the amount of fine-tuning  (see equation
(\ref{dktc})) in $\Delta\kappa_{tc}$ required to satisfy
the bound  $\Delta \rho_* < 0.4$\%.  The region excluded by the
experimental constraint on $\Delta \rho_*$ is above the solid curve.

\end{document}